\documentclass[superscriptaddress,onecolumn]{revtex4}
\usepackage{graphicx}
\usepackage{amssymb}
\usepackage{color}

\begin{document}

\title{Dilepton flow and deconfinement phase transition in heavy ion collisions}

\author{Jian Deng}

\affiliation{Interdisciplinary Center for Theoretical Study and Department of
Modern Physics, University of Science and Technology of China, Anhui
230026, People's Republic of China}
\affiliation{School of Physics, Shandong University, Jinan, Shandong 250100, China}

\author{Qun Wang}

\affiliation{Interdisciplinary Center for Theoretical Study and Department of
Modern Physics, University of Science and Technology of China, Anhui
230026, People's Republic of China}

\author{Nu Xu}

\affiliation{Nuclear Science Division, Lawrence Berkeley National Laboratory,
Berkeley, California 94720, USA}

\author{Pengfei Zhuang}

\affiliation{Physics Department, Tsinghua University, Beijing 100084, China}

\begin{abstract}
The dilepton radial flow in Au+Au collisions at $\sqrt{s_{NN}}=$200
GeV is investigated. The space-time evolution of the fireball is
described by a $2+1$ dimensional ideal hydrodynamics with a variety
of equations of state. The slope parameters of the transverse
momentum spectra from the partonic and hadronic phases show distinct
features and are sensitive to equation of state parameters. The
elliptic flow and breaking of $M_T$ scaling are also studied and
have distinct features for the two phases. These features can serve
as clean signals for the formation of a quark-gluon plasma in
ultra-relativistic heavy ion collisions.
\end{abstract}

\maketitle

\section{Introduction}
Among all observables for determining the quark gluon plasma (QGP)
in heavy ion collisions, the electromagnetic probes such as photons
and dileptons are expected to provide clean signitures due to their
instant emissions once produced \cite{McLerran:1984ay,Kajantie:1986dh,
Rapp:1995zy,Rapp:1997fs,Rapp:1999qu,Rapp:1999ej,
Alam:1999sc,vanHees:2007th,Chatterjee:2007xk,Dusling:2008xj,Mohanty:2010sw}.
These thermal photons and dileptons contain undistorted information
about the space-time trace of the new state of matter formed in such
collisions. There are many sources of dileptons in heavy ion collisions.
In the lower invariant mass region ($M\lesssim$1 GeV) dileptons are
mainly from resonance decays and may be related to chiral symmetry
restoration \cite{Pisarski:1981mq,Arnaldi:2006jq,vanHees:2006ng,Ruppert:2007cr}.
In the higher invariant mass region ($M\gtrsim3$ GeV) dileptons are
dominated by the Drell-Yan process and charmonium decays. For moderate
invariant mass dileptons ($1\lesssim M\lesssim3$ GeV), it was argued
that dileptons from semileptonic decays of correlated open charm in
$pp$ collisions are dominant \cite{PHENIX:2008asa}. But in Au+Au
collisions, both charm related single lepton contributions and their
dynamic correlations are expected to be suppressed by medium modification.
Therefore thermal radiation may play important role in the intermediate
mass region and the dilepton spectra can be used to extract thermodynamic
parameters of the fireball.

The observation of jet quenching and strong elliptic flow at the
Relativistic Heavy Ion Collider (RHIC) at Brookhaven National
Laboratory tell us that the dense matter produced at RHIC interacts
strongly and may reach local thermalization in very short time
\cite{Ackermann:2000tr,Adcox:2002ms}, implying that ideal
hydrodynamic models are applicable to such systems
\cite{Shuryak:2004cy,gyulassy:2005}. In contrast to hadronic flow,
the observables of dileptons are more direct and penetrating probes
to the early space-time profile of the QGP
\cite{Chatterjee:2007xk,Dusling:2008xj}. The radial flow of thermal
dileptons has been measured at CERN super-proton synchrotron (SPS)
by the NA60 collaboration \cite{Arnaldi:2007ru}. It was found that
the inverse slope parameter $T_{eff}$ increases with the invariant
mass $M$ of the lepton pair below the $\rho$ meson mass and then
starts decreasing above $M\sim1$ \cite{Arnaldi:2007ru}. The reason
for the drop of $T_{eff}$ around $M\sim1$ GeV is not fully
understood although it was thought to be an indication of the transition
to an emission source with much smaller flow possibly a partonic source  
\cite{Arnaldi:2007ru,Renk:2006qr}. The observed strong correlation
of $T_{eff}$ versus $M$ in the region $M\lesssim1$ GeV is mainly due
to the collectivity developed in the hadronic stage at SPS energies.
Should the collectivity have been developed in the partonic phase of
the evolution in high energy nuclear collisions, RHIC and/or LHC,
one would also expect to see the increase of $T_{eff}$ in the
intermediate mass region $1\lesssim M\lesssim3$ GeV. Inspired by the
NA60 result and in order to develope clean observables for the
formation of the QGP, we propose to study the transverse momentum
distributions of di-electrons over the entire region of $0.5\lesssim
M\lesssim 3$ GeV.

\begin{figure}
\caption{\label{fig:eos}Equations of state: QGP-EOS (upper-left),
HG-EOS (upper-right), MIX-EOS (lower-left), LAT-EOS (lower-right). }
\includegraphics[scale=0.35]{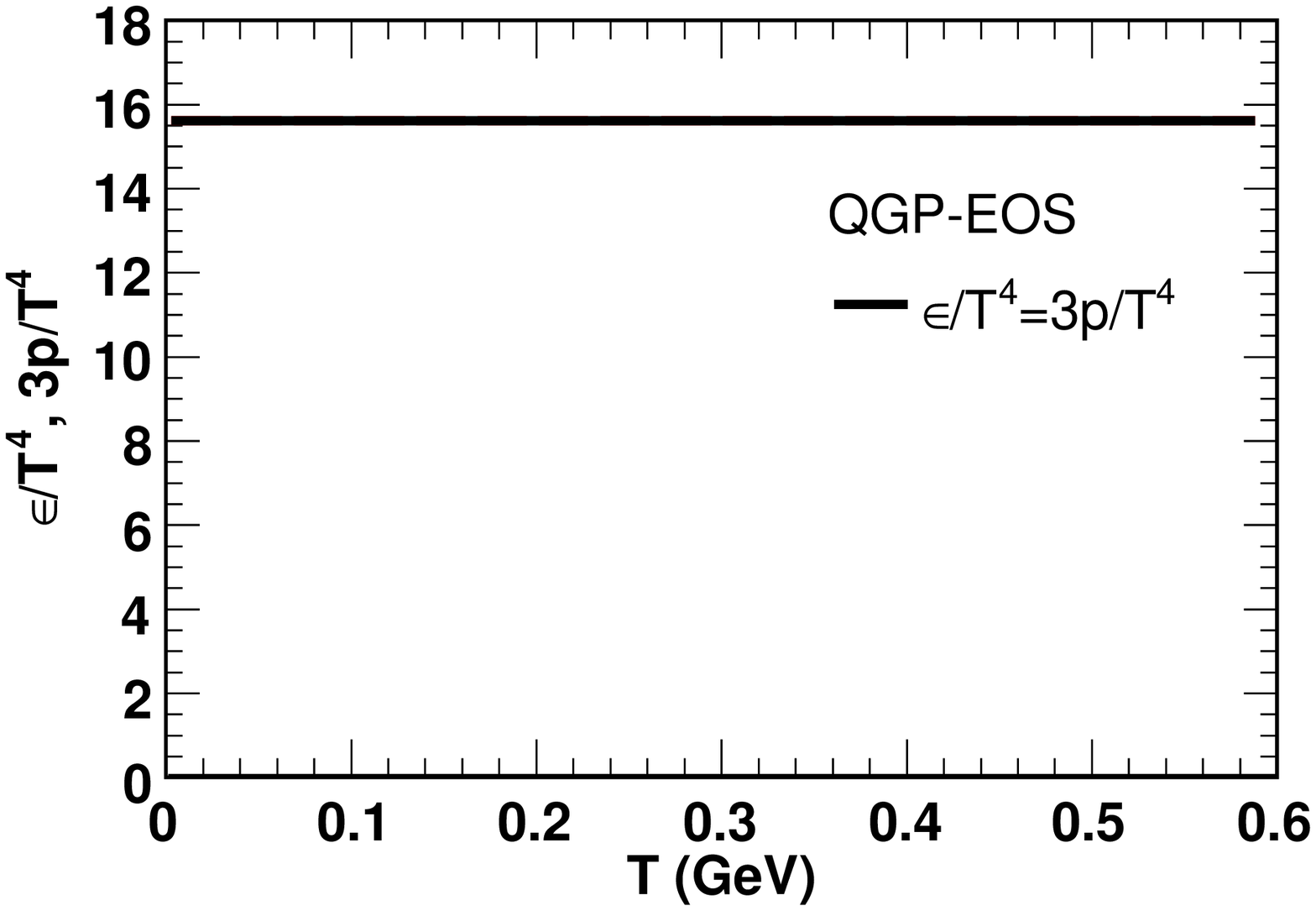}\includegraphics[scale=0.35]{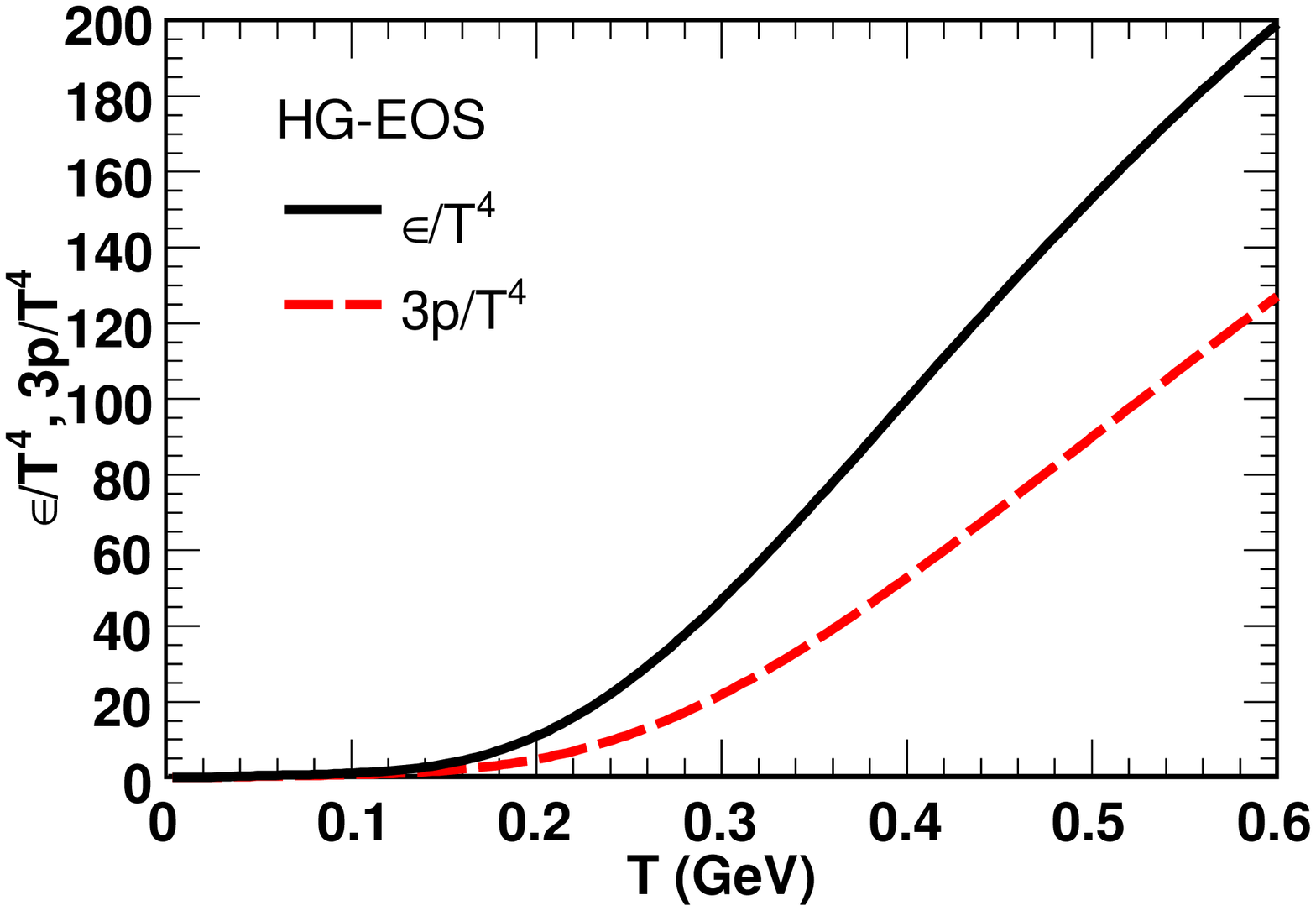}
\includegraphics[scale=0.35]{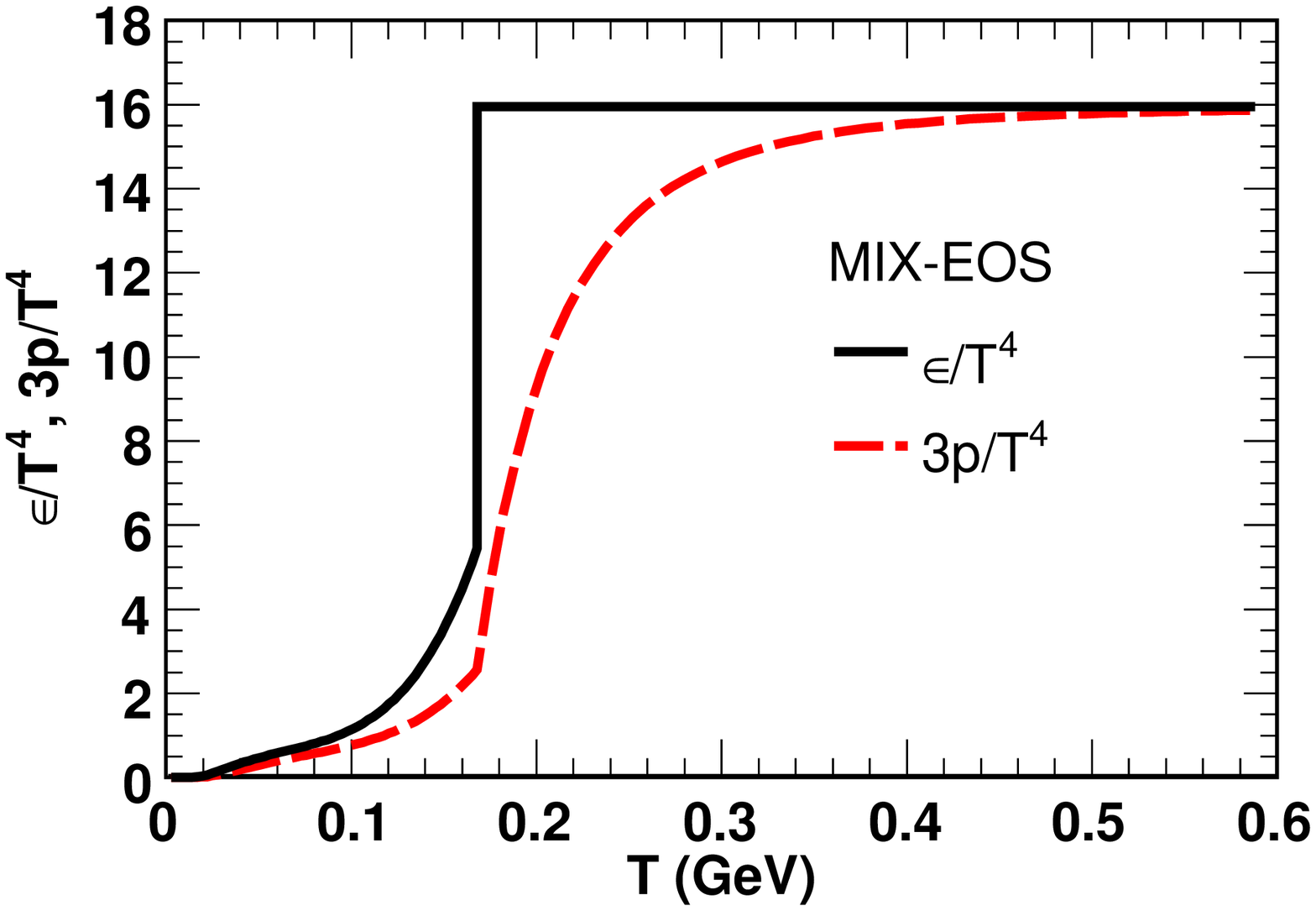}\includegraphics[scale=0.35]{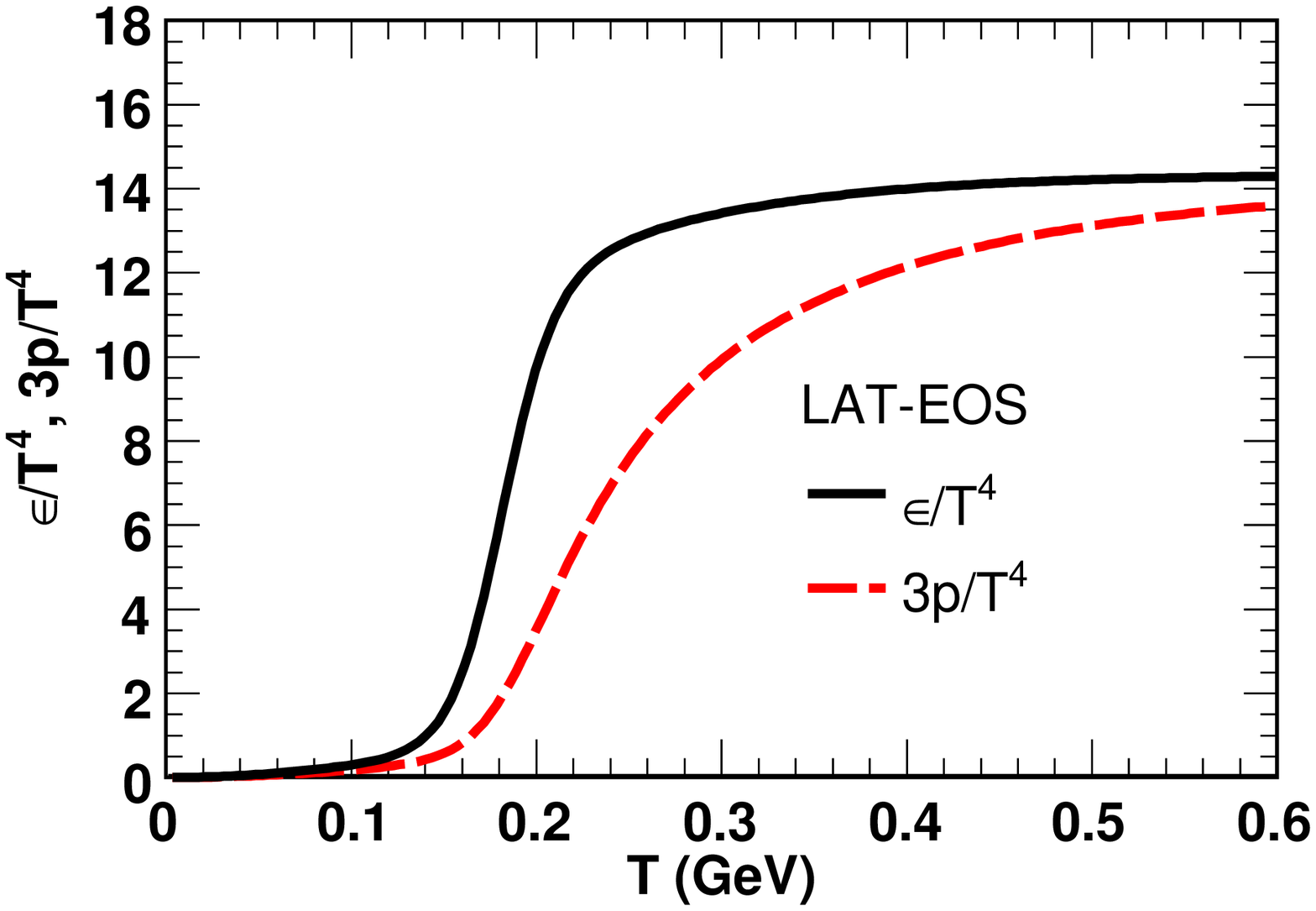}
\end{figure}

In this paper we use a 2+1 dimensional ideal hydrodynamic model to
give the space-time evolution of the medium created in Au+Au
collisions at $\sqrt{s_{NN}}=$200 GeV. Our program gives results
consistent with AZHYDRO \cite{Kolb:2000sd}. To include the
pre-equilibrium emission of the di-electrons at very beginning, we
set the initial time for the hydrodynamic evolution $\tau_{0}=0.2$
fm/c as in Ref. \cite{Chatterjee:2007xk} instead of $\tau_{0}=0.6$
fm/c in previous studies. The initial transverse energy density is
calculated in the Glauber model with the peak temperature being at
about 520 MeV in central collisions. Another critical input is the
equations of state (EOS) of the dense matter \cite{Teaney:2000cw}.
We choose four types of EOS, (i) QGP-EOS, the ideal gas EOS for
3-flavor QGP, $\epsilon=3p=(19/12)\pi^{2}T^{4}$ without the hadronic
phase; (ii) HG-EOS, the resonance hadron gas EOS for the hadron gas
(HG) \cite{Sollfrank:1996hd} without the partonic phase; (iii)
MIX-EOS, the one with the first order phase transition with both the
partonic and hadronic phases \cite{Sollfrank:1996hd}; (iv) LAT-EOS,
the one extracted from the lattice calculations
\cite{Bazavov:2009zn}. Note that in the MIX-EOS and LAT-EOS there
are partonic and hadronic components. These EOS are shown in Fig.
\ref{fig:eos}. We put emphasis on the lattice EOS which can describe
the crossover from resonance HG to QGP in the temperature range
180-200 MeV. Other EOS are also considered in our calculations for
the purpose of comparison. Althrough the crossover does not have a
rigorous critical temperature $T_{c}$ to separate the QGP from the
HG phase, we still choose $T_{c}=180$ MeV as a tuning parameter. The
freezeout temperature is set to $T_{f}=120$ MeV below which there is
no di-electron emission from the HG sector. The post-freezeout
di-electron decay of the $\rho$ mesons is not included directly in
our calculation, whose effect on the di-electron spectra can be
partly taken into account by lower the freezeout temperature. The
fine tuning of $T_{c}$ and $T_{f}$ does not qualitatively change our
results and conclusions.

\section{Dilepton Yields} 

In the thermalized medium HG or QGP, the rate for the di-lepton production 
per unit volume is given by \cite{Rapp:1995zy,Rapp:1997fs,Rapp:1999qu}, 
\begin{eqnarray}
\frac{d^{4}N}{d^{4}xd^{4}p} & = & 
-\frac{\alpha}{4\pi^{4}}\frac{1}{M^{2}}n_{B}(p \cdot u)
\left(1+\frac{2m_{l}^{2}}{M^{2}}\right)\sqrt{1-\frac{4m_{l}^{2}}{M^{2}}}
\mathrm{Im}\Pi^R (p).
\end{eqnarray}
Here $m_l$ is the lepton mass, $\alpha = e^2/(4\pi)$ is the fine structure constant 
with the electric charge $e$ for leptons, $p=(p_0,\mathbf{p})=p_1+p_2$  
is the dilepton 4-momentum and $M=\sqrt{p^2}$, 
$n_B(p\cdot u)=1/(e^{p\cdot u/T}-1)$ is the Bose distribution function 
with $T$ and $u$ the local temperature and medium velocity respectively, 
$\Pi^R_{\mu\nu}$ is the retarded photon polarization tensor 
from the quark-loop or the hadronic loop, and $\Pi^R = \frac 13 \Pi_{\mu}^{R\mu}$. 
For the hadron gas, $\Pi^R$ is further related to the retarded rho-meson 
propagator $D_{\rho}^{R}$ via $\mathrm{Im}\Pi^{R}=-(e^{2}m_{\rho}^{4}/g_{\rho}^{2})\mathrm{Im}D_{\rho}^{R}$, 
where $g_\rho$ is the photon-rho-meson coupling constant 
in the vector meson dominance model, and $m_\rho$ is the rho-meson mass. 
The imaginary part of the retarded $\rho$-meson propagator is given by, 
\begin{equation}
\mathrm{Im} D_{\rho}^{R}=\frac{\mathrm{Im}\Pi_{\rho}^{R}}{(p^2-m_{\rho}^{2}+\mathrm{Re}
\Pi_{\rho}^{R})^{2}+(\mathrm{Im}\Pi_{\rho}^{R})^{2}} ,
\label{eq:d-rho}
\end{equation}
where $\Pi_{\rho}^{R}$ is the retarded rho-meson polarization tensor contraction. 
We assume that the hadron gas is mainly composed of mesons and consider 
following mesonic vertices for the $\rho$ meson \cite{Rapp:1999qu} in $\Pi_{\rho}^{R}$: 
$\rho \pi X$ and $\rho K K'(1270)$, where $X$ denotes meson resonances below 1300 MeV, 
namely, $\omega$, $h_1 (1170)$, $a_1(1260)$ and $\pi' (1300)$. 
Here we neglect the baryonic contributions since the collisional energy considered in this paper 
is the RHIC energy (200 GeV) where the baryon yields are smaller than meson yields.
The baryonic contributions may broaden the width of the rho-meson even more. 
But we argue that they will not qualitatively change the behavior of $T_{eff}$ 
in the mass range considered in this paper, considering that the mesonic resonance 
contributions do not qualitatively change the dilepton spectra and then the behavior of 
$T_{eff}$ compared to the dominant $\pi\pi$ process in this mass range. 
Note that $\Pi_{\rho}^{R}$ depends on thermal distributions of mesons 
and then on local temperature $T(x)$ and medium velocity $u(x)$ 
determined by hydrodynamic simulation, which are functions of space-time position. 
We only take into account the processes mediated through the $\rho$ meson, because the
di-electron production is dominated by the isovector channel instead
of the isoscalar one. According to the flavor $SU(3)$ quark model,
the relative weight of the electromagnetic couplings for vector
mesons $V=\rho,\omega,\phi$ is about 9:1:2, roughly consistent to
the electromagnetic decay widths $\Gamma_{V\rightarrow
ee}=7.0,0.6,1.27$ KeV respectively \cite{vanHees:2007th}. 
On the other hand, $\omega$ and $\phi$ mesons can be easily identified 
in experiments. We also neglect the Daliz decay channels for $\eta$ and $\pi^{0}$:
$\eta\rightarrow e^{+}e^{-}\gamma$, $\mu^{+}\mu^{-}\gamma$ and
$\pi^{0}\rightarrow e^{+}e^{-}\gamma$. The pion spectrum is mainly
below $m_{\pi}=135$ MeV and irrelevant to our current range of $M$.
The $\eta$ contribution can be easily deducted as the background in
experiments due to its very long lifetime (about $1.5\times10^{5}$
fm/c), leading to decays outside the freezeout scope.

After making approximation $n_B(p\cdot u)\approx e^{p\cdot u/T}$ and 
then integrating over the lepton pair rapidity $y$, 
we obtain the differential cross section for an expanding fireball \cite{Kajantie:1986cu},
\begin{eqnarray}
\frac{d^{4}N}{p_{T}dp_{T}MdMd\phi_{p}} 
& \approx & -\frac{\alpha}{2\pi^4}\frac{1}{M^{2}}
\left(1+\frac{2m^{2}}{M^{2}}\right)\sqrt{1-\frac{4m^{2}}{M^{2}}} \nonumber \\
 &  & \times\int d^{4}x \exp\left[\frac{1}{T}\gamma_{T}M_{T}v_{T}p_{T}\cos(\phi_{v}-\phi_{p})\right]
K_{0}\left(\frac{\gamma_{T}M_{T}}{T}\right) \mathrm{Im}\Pi ^{R} , 
\label{eq:dndptdm02} 
\end{eqnarray}
where $p_{T}\equiv|\mathbf{p}_{T}|$ is the scalar transverse
momentum of the lepton pair, $\phi_{p}$ and $\phi_{v}$ are the
azimuthal angles of $\mathbf{p}_{T}$ and the local transverse fluid
velocity $\mathbf{v}_{T}$ respectively,
$\gamma_{T}=1/\sqrt{1-v_{T}^{2}}$ is the local transverse Lorentz
factor for the fluid element, and $K_{0}$ is the modified Bessel
function of the second kind. Note that the encoding of space-time
history of the fireball is realized by integrals over fluid
coordinates $d^{4}x=\tau d\tau d\eta d^{2}\mathbf{x}_{T}$, where
$\tau,\eta$ are proper time and space-time rapidity respectively and
$\mathbf{x}_{T}$ is the transverse position of the fluid element.
For central collisions with azimuthal symmetry, the angular integral
can be worked out by $\int d\phi\exp[x\cos\phi]=2\pi I_{0}(x)$
analytically with $I_{0}$ being the modified Bessel function of the
first kind.

\begin{figure}
\caption{\label{fig:mult}(Color online) Differential multiplicity as
a function of proper time(left panel) and the di-electron invariant
mass (right panel). The LAT-EOS \cite{Bazavov:2009zn} is used in the
calculation and the contributions from the QGP and HG are shown in
dashed and dotted lines, respectively. The dashed-dotted-dotted line 
is the HG contribution from the $\rho\pi\pi$ vertex only. 
The contribution from the HG is dominated by $\rho$ mesons. } 
\includegraphics[scale=0.4]{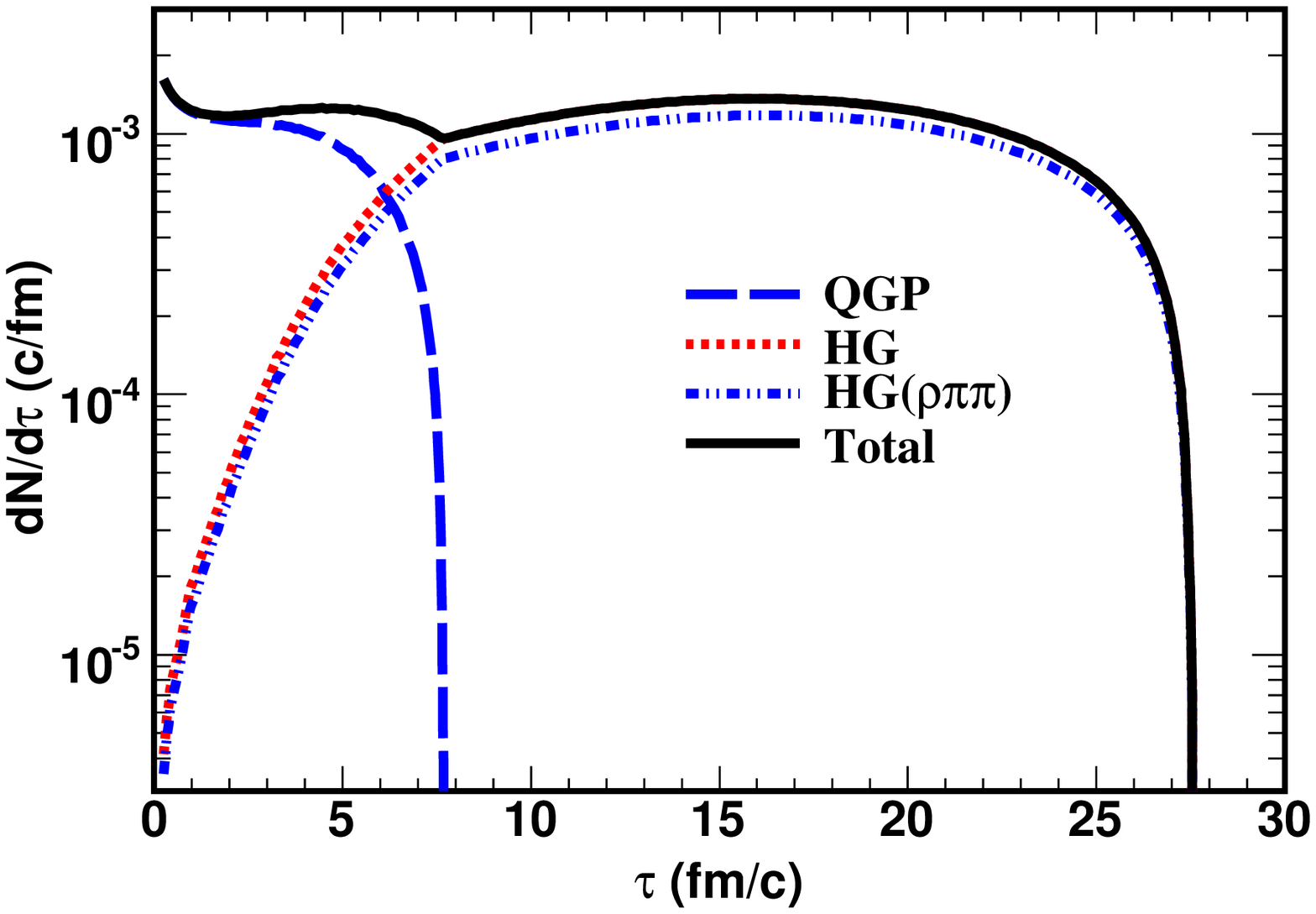} \includegraphics[scale=0.4]{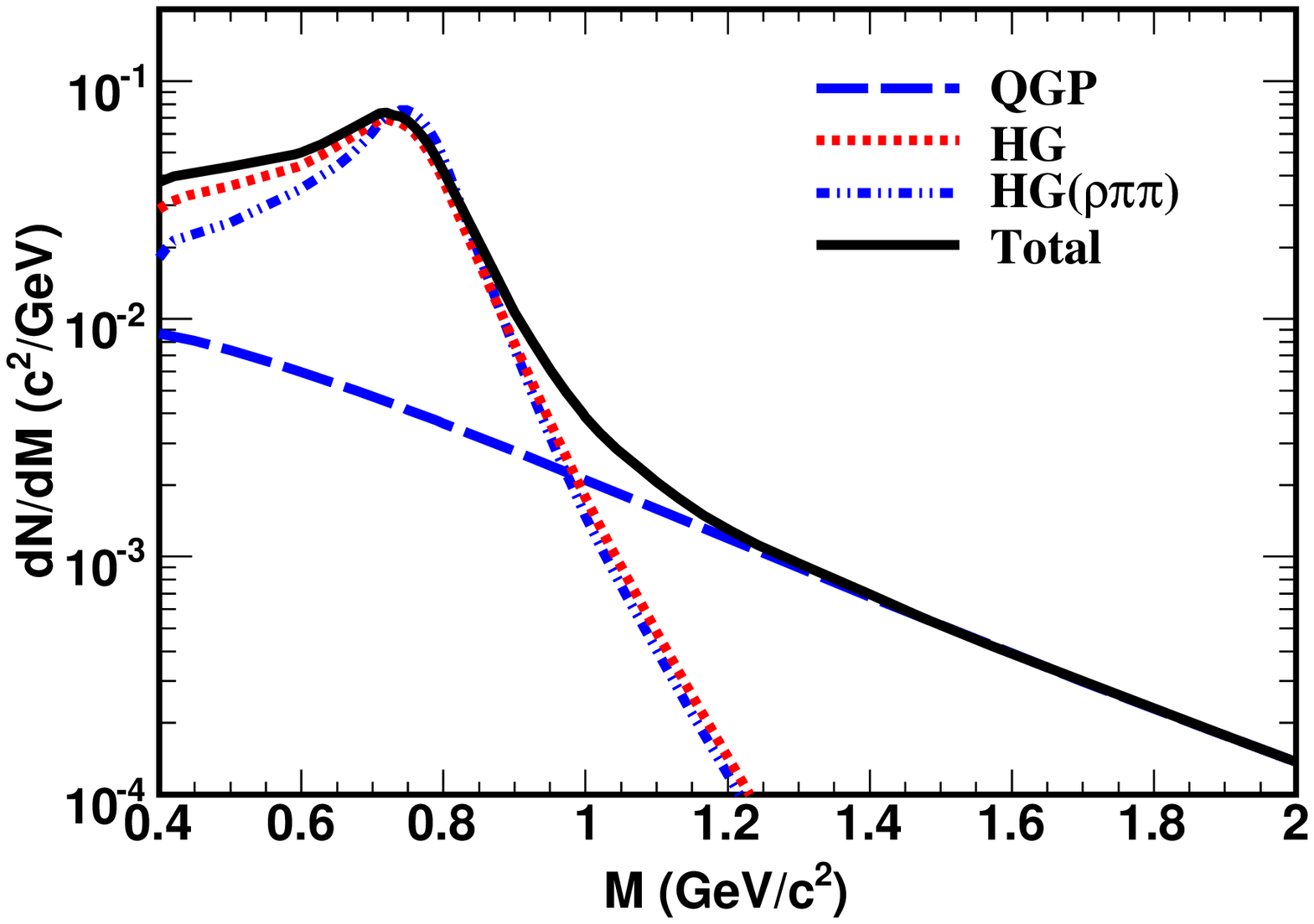}
\end{figure}

The evolution of the rate with the proper time $\tau$ is shown in
the left panel of Fig. \ref{fig:mult}. The emission rate is
proportional to $T^{4}S\tau$ where $S$ is the transverse area. For
the harder EOS like the QGP ideal gas there is a rapid expansion
leading to a very fast decreasing of $T$ and $S$. As a result the
QGP freezes out in a very short time with less dilepton emission.
But with softer EOS like the lattice one, $T$ decreases slowly as
$S$ expands, making the freezeout isothermal lines expand and remain
almost constant at large radii, which greatly increases the
di-electron emission in the HG phase.

The numerical results for the invariant mass spectra of the
multiplicity with the LAT-EOS in central collisions are shown in the
right panel of Fig. \ref{fig:mult}. The contributions from the QGP
and the HG components are distinguished. The shapes of the spectra
with the QGP-EOS only or the HG-EOS only are similar to the QGP or
HG component (blue-dashed/red-dotted line) here. One can see that
the QGP/HG contribution dominates in the region $M\gtrless1$ GeV.
The emission from the QGP phase in the early stage with high
temperatures contributes to the hard parts of the spectra with large
$M$ and $m_{T}$. The HG phase in later stage with low temperature
contributes in small $M$ region and is shaped by the $\rho$ meson
form factor, but a sizable collective flow developed in this stage
can harden the transverse momentum spectra \cite{Kajantie:1986cu}.
Also we have shown the HG contribution from 
the $\rho\pi\pi$ vertex only, which is lower than the full HG contribution 
(with all resonances below 1300 MeV included) in the lower mass region. 
This indicates the broadening effect to the $\rho$ meson spectra 
from the resonances' contribution.

\section{Effective Temperature, Elliptic Flow and $M_T$ scaling}

To discuss the sensitivity of dilepton signals to different stages
and collectivity developed in the history of the fireball, it is
valuable to investigate dilepton transverse momentum spectra and its
inverse slope parameter. We can use the reduced transverse mass
$m_{T}\equiv M_{T}-M$ with $M_{T}=\sqrt{M^{2}+p_{T}^{2}}$ to replace
$p_{T}$ as the variable. After carrying out the space-time integrals
we assume that the transverse spectra can be approximately
parameterized as \cite{Schnedermann:1993ws,Gorenstein:2001ti},
\begin{eqnarray}
\frac{d^{2}N}{m_{T}dm_{T}MdM} & \sim & \sqrt{\frac{\overline{T}}{\overline{\gamma}_{T}}}
\frac{\sqrt{m_{T}+M}}{m_{T}}\exp\left(-\frac{m_{T}+M}{T_{eff}}\right),\label{eq:teff01}\end{eqnarray}
with the average temperature $\overline{T}$, the average transverse
velocity $\overline{v}_{T}$ of the fluid and its Lorentz factor $\overline{\gamma}_{T}=1/\sqrt{1-\overline{v}_{T}^{2}}$.
The asymptotic forms of the slope parameter $T_{eff}$ can be written as, 
\begin{equation}
T_{eff}\sim\left\{ \begin{array}{c}
\overline{T}+M^{*}\overline{v}_{T}^{2},\quad\mathrm{for}\; p_{T}\ll M\\
\overline{T}\sqrt{\frac{1+\overline{v}_{T}}{1-\overline{v}_{T}}},\quad\mathrm{for}\;
p_{T}\gg M\end{array}\right.,
\label{eq:teff02}
\end{equation} 
where $M^{*}$ is a monotonic function of $M$. As emissions in different
space-time are encoded in the spectra (\ref{eq:teff01}), the slope
parameter $T_{eff}$ depends on the fitting window of $M$ and
$p_{T}$, so do $\overline{T}$ and $\overline{v}_{T}$.

\begin{figure}
\caption{\label{fig:teff}(Color online) Left panel: slope parameter
$T_{eff}$ as a function of $M$ with the LAT-EOS and for three
$m_{T}$ values. Results for three $m_{T}$ values are shown as solid,
dashed and dotted lines. For $m_{T}=0.5$ GeV, thin red and blue
lines with circles and triangles represent the results
extracted from contributions of the HG and QGP phases, respectively.
Right panel: The slope parameter $T_{eff}$ calculated for four
equations of state at fixed $m_{T}=2.5$ GeV. }
\includegraphics[scale=0.4]{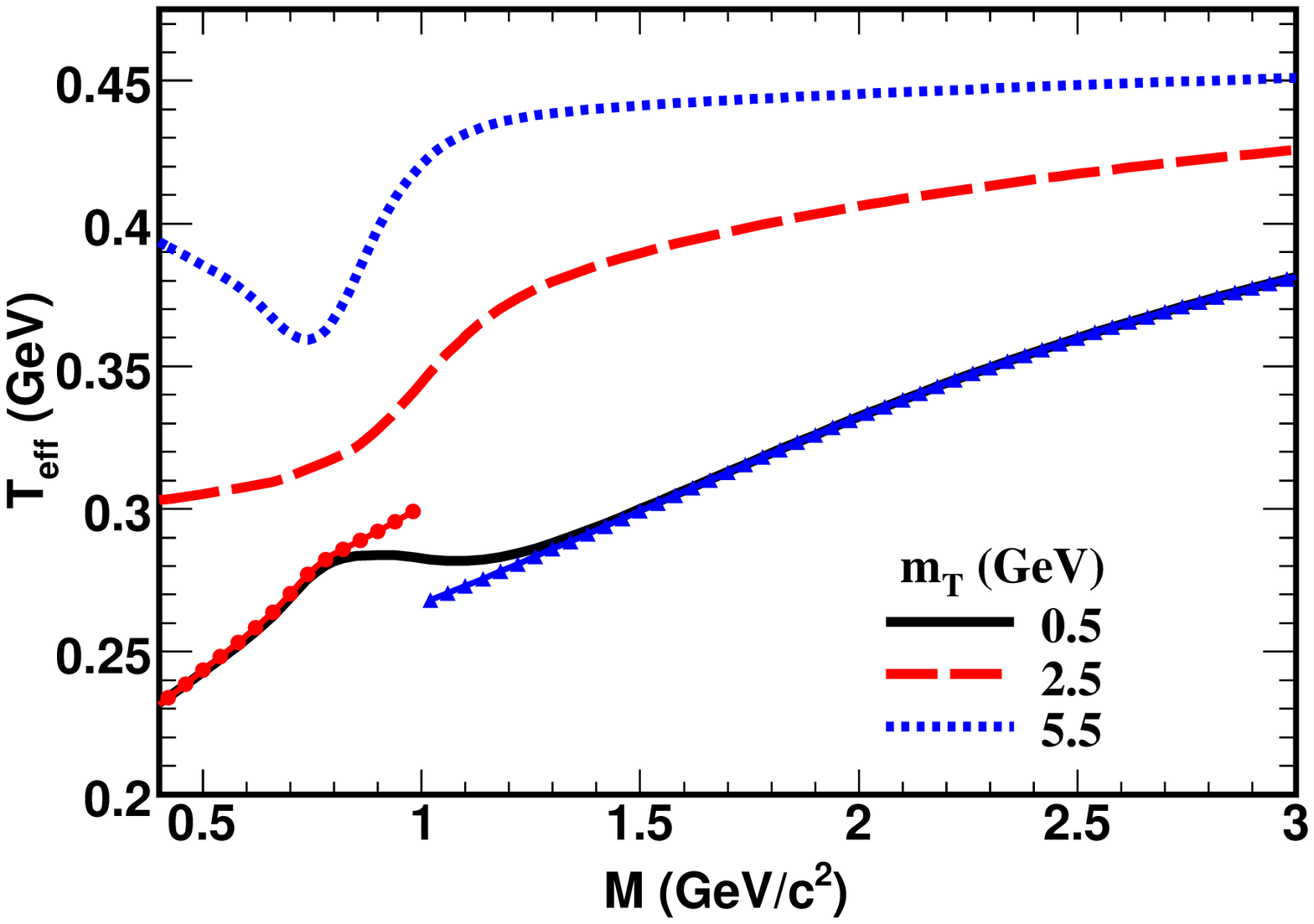}\includegraphics[scale=0.4]{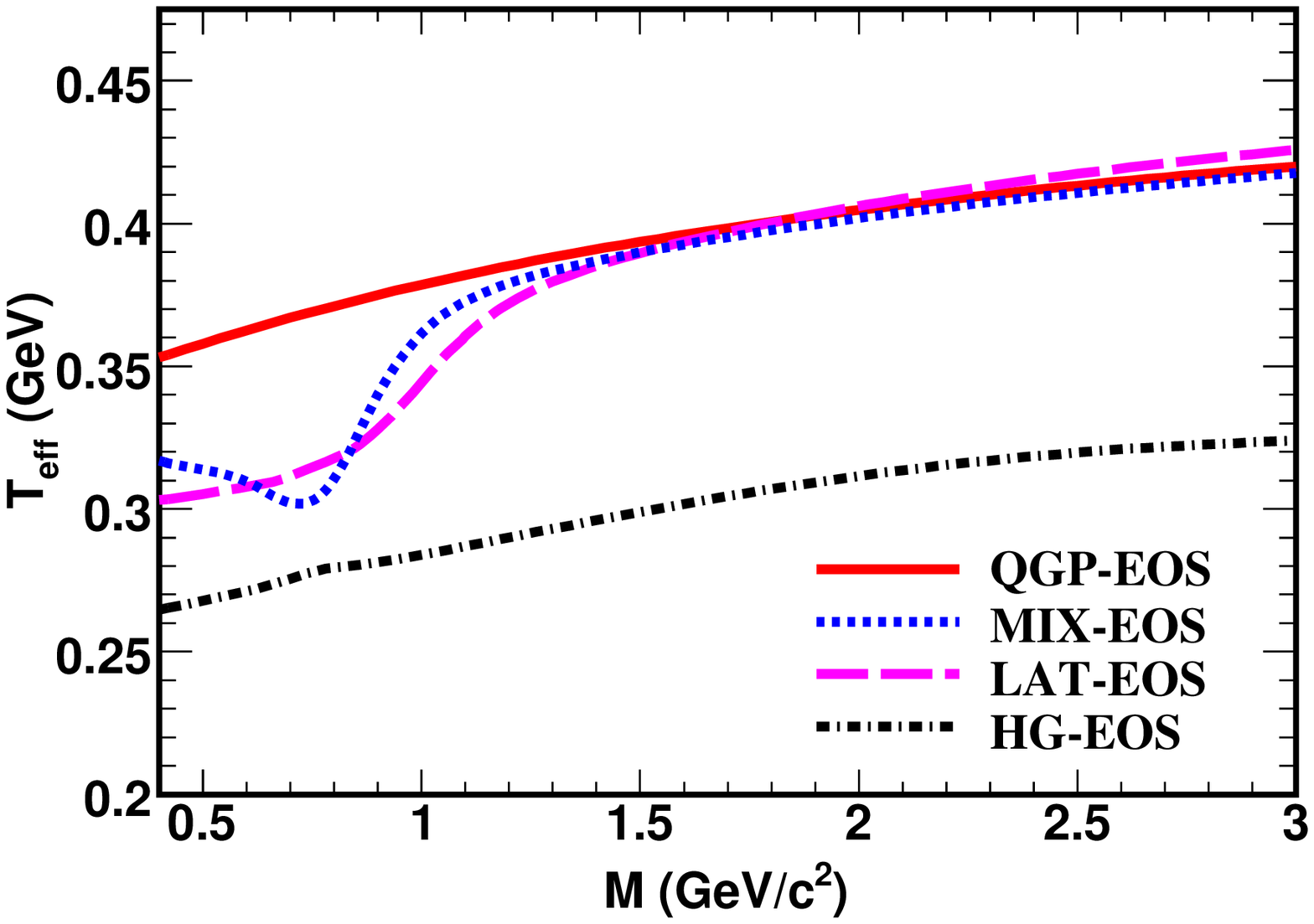}
\end{figure}

The slope parameters $T_{eff}$ versus $M$ from the LAT-EOS are shown
in the left panel of Fig. \ref{fig:teff}. For low $p_{T}$, e.g.
$m_{T}=0.5$ GeV, $T_{eff}$ approximately follows
$T_{eff}\sim\overline{T}+M^{*}\overline{v}_{T}^{2}$ as in Eq.
(\ref{eq:teff02}). The curves below 1 GeV have smaller
$\overline{T}$ and larger $\overline{v}_{T}$, while above 1 GeV they
have an opposite trend. For even higher $M$ the increase of
$T_{eff}$ is not caused by the collective flow since larger $M$
reflects the higher temperature in earlier emission when the
collective flow has not fully developed. The disconnected curves
with solid circiles and triangles are $T_{eff}$ extracted from the contributions
of the HG or QGP phase only. They refl{}ect different trends of
$T_{eff}$ below/above 1 GeV from the HG/QGP phase. When $m_{T}$ is
larger, e.g. $m_{T}=2.5$ GeV, $T_{eff}$ does not follow the above
formula for low $p_{T}$, but the same trends still exist. For very
large $m_{T}$, e.g. $m_{T}=5.5$ GeV, $T_{eff}$ follows the
blue-shift formula $T_{eff}\sim
T\sqrt{\frac{1+\overline{v}_{T}}{1-\overline{v}_{T}}}$, which is
independent of $M$. The valley at $M\sim m_{\rho}$ is due to the
fact that the QGP component (with higher $T_{eff}$) is dominant over
the HG one (with lower $T_{eff}$) except in the region near $M\sim
m_{\rho}$. One can see in the figure that $T_{eff}$ increases with
$m_{T}$, since the larger $m_{T}$ probe the earlier state of the
fireball with high temperatures. For the MIX-EOS, since there is a
large contribution from the coexisting stage of two phases with the
same temperature and fluid velocity, the difference in $T_{eff}$
between the QGP and HG phases is smaller than the case of the
LAT-EOS. For other two types of EOS, the QGP and HG ones, $T_{eff}$
simply increase with $M$ monotonously. In right panel of Fig.
\ref{fig:teff}, we compare $T_{eff}$ with different EOS. It is found
that the magnitude of $T_{eff}$ for the HG-EOS is much smaller than
that for other EOS with partonic phase. Such behaviors of
di-electron $T_{eff}$ are expected to be measured and tested in the
future experiments.

\begin{figure}
\caption{\label{fig:Tctau0effect} Slope parameter $T_{eff}$ as a
function of $M$ with the dependence on the initial time $\tau_{0}$
(left panel) and on the phase transition temperature $T_{c}$ (right
panel). We choose $m_T=0.5$ GeV and use LAT-EOS for both panels. 
The red and blue lines with circles represent the results
extracted from contributions of the HG and QGP phases, respectively.}
\includegraphics[scale=0.4]{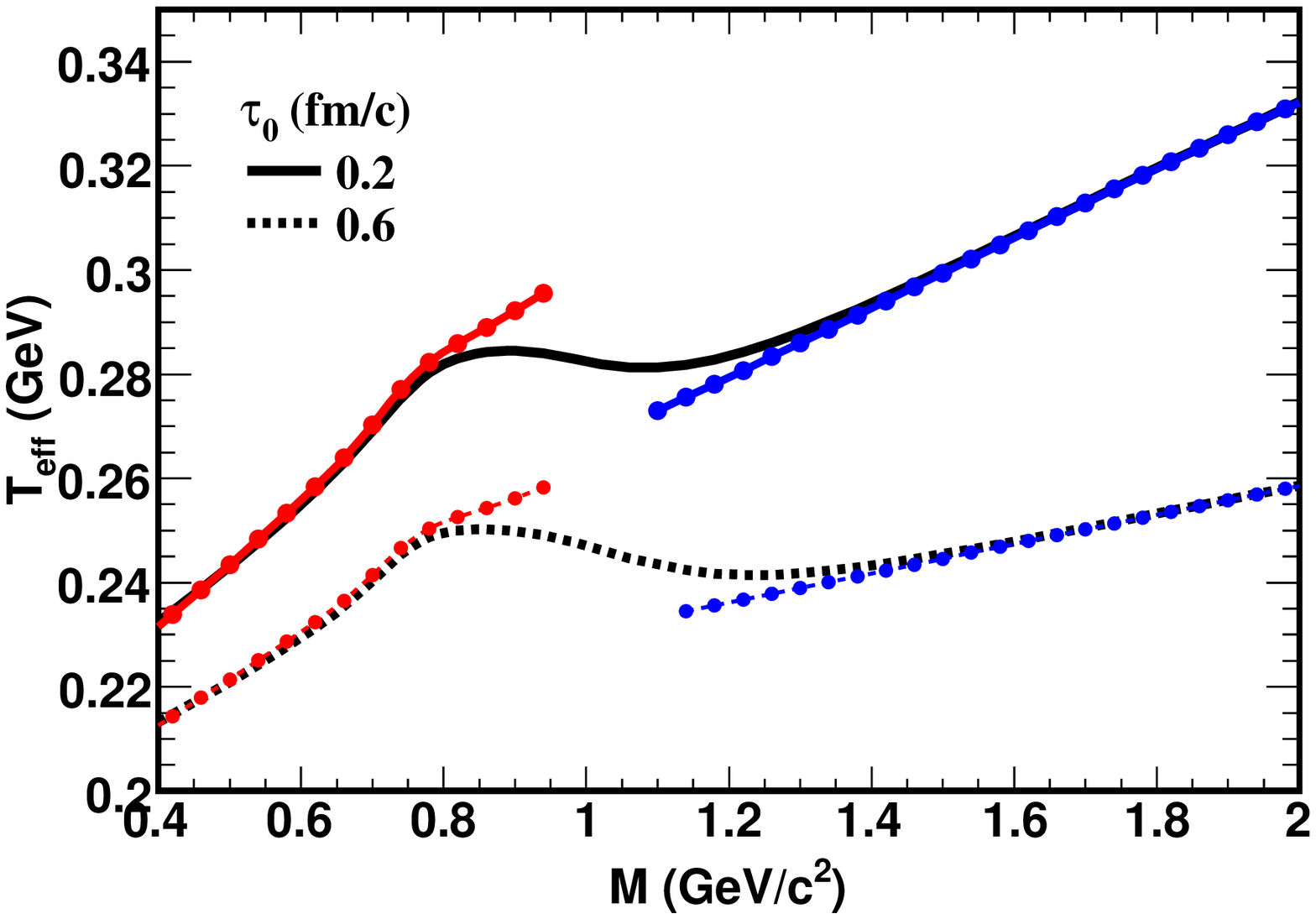}
\includegraphics[scale=0.4]{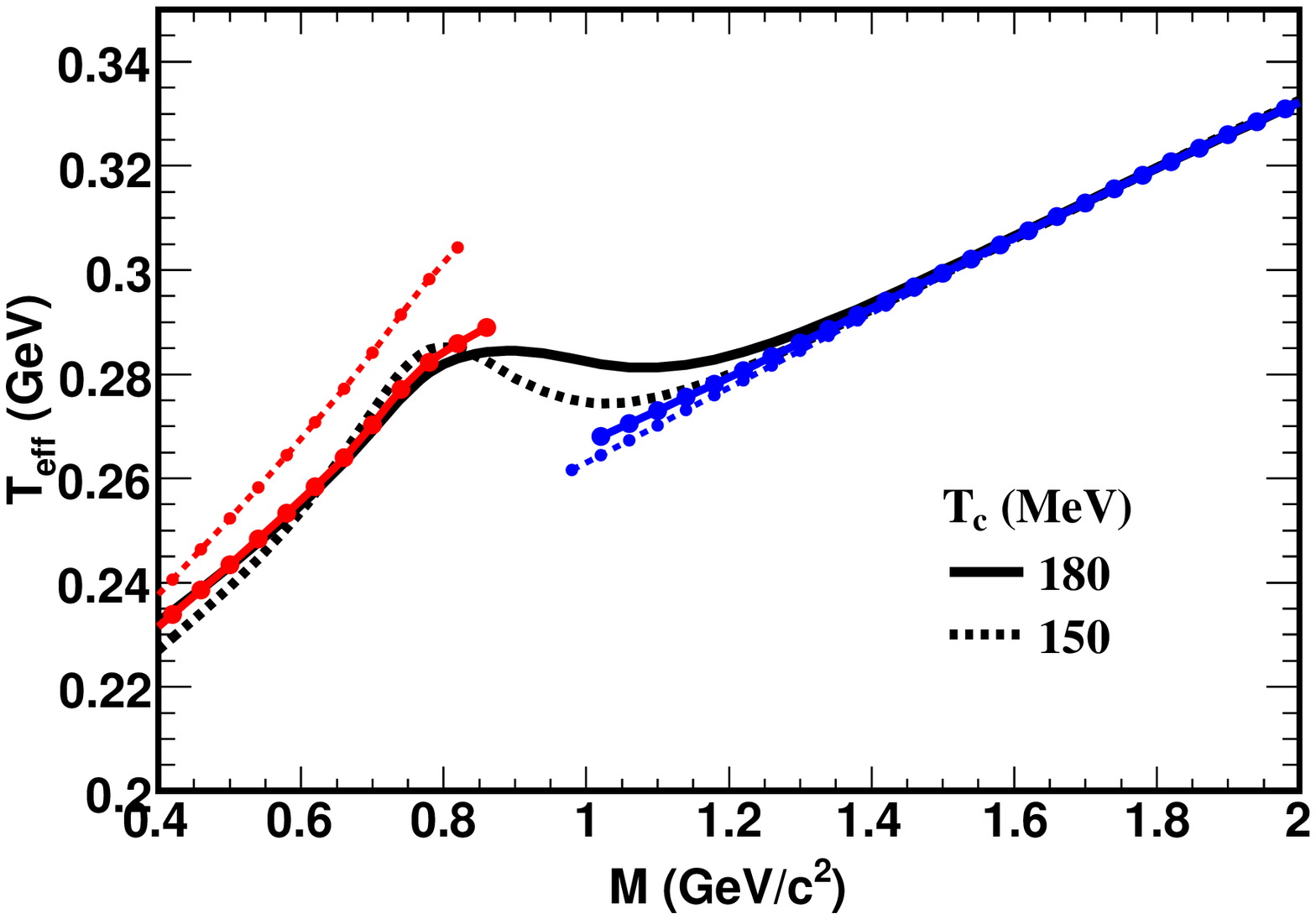}
\end{figure}

In Fig. \ref{fig:Tctau0effect} we show the dependence of $T_{eff}$
on the initial time $\tau_{0}$ for hydrodynamic evolution and the
phase transition temperature $T_{c}$. When we tune $\tau_{0}$ from
0.2 fm/c to 0.6 fm/c and change the initial peak temperature from
520 MeV to 350 MeV, or in other words, delay the hydrodynamic
process, the magnitudes of $T_{eff}$, $\overline{T}$ and
$\overline{v}_{T}$ decrease due to smaller initial temperature and
acceleration but their structure remains the same. When changing
$T_{c}$ from 180 to 150 MeV, the magnitude of $T_{eff}$ does not
change much, but the slope increases in the small $M$ region. This
can be understood because a smaller transition temperature means
that the HG phase occupies outer layer of the fireball with larger
radial flow velocities.

Now we can compare our result for $T_{eff}$ with NA60 data \cite{Arnaldi:2007ru}. 
The $T_{eff}$ curve of NA60 is extracted from 
low $m_T$ spectra. It is comparable to our result 
in Fig. \ref{fig:Tctau0effect} with $m_T=0.5$ GeV. 
But the trend above $M=1$ GeV in NA60 data is different: it is downward 
while our curve is upward after a transition region between 0.8 and 1.2 GeV. 
We now give the reason for such a difference. 
We note that the region of $M>1$ GeV probes earlier stage of the fireball, 
where the QGP emission dominates over the hadronic emission. 
Such a different trend is just the manifestation of the radial flow and average temperature effects 
of the QGP phase. At the NA60 energy, the life of the QGP phase 
(if there is any) is much  shorter and its average temperature is lower 
(near freeze-out $T_f$), the radial flow does not have enough time 
to develope and the average temperature is just above $T_f$. 
For $M<1$ GeV, the dominant contribution is from the hadronic phase 
which appears at later stage of the evolution and has larger collective flow, 
so $T_{eff}$ is larger since it has a component proportional to 
the flow velocity squared times an effective mass, 
see first line of Eq. (\ref{eq:teff02}). That is the reason for the upward/downward trend 
below/above 1 GeV at the NA60 energy. In contrast, at the RHIC energy, 
the life of the QGP phase is longer and there is some time for the radial flow 
to develope and make $T_{eff}$ larger. Another effect is that the average temperature 
in the QGP phase is larger than at the NA60 energy. This the reason for 
the upward trend above 1 GeV in our result.

\begin{figure}
\caption{\label{fig:Mtscaling}Elliptic flow $v_2$ (left panel) and
the ratio $R$ for the $M_T$ scaling (right panel) for different EOS. 
For $v_2$ the impact parameter is set to $b=7$ fm. 
For $R$ we scan all values of the impact parameter, i.e. $b=[0,15]$ fm.}
\includegraphics[scale=0.35]{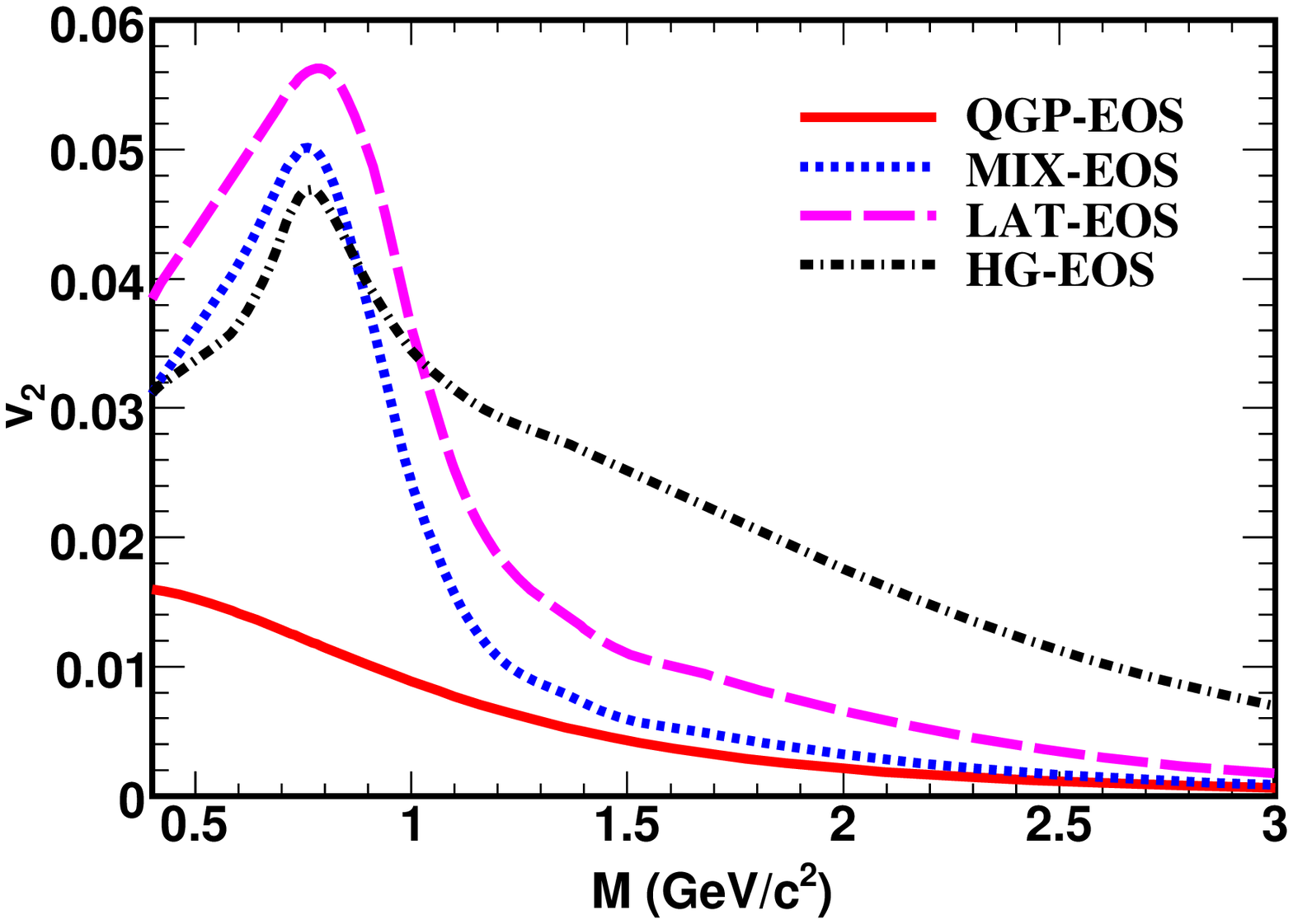}\includegraphics[scale=0.35]{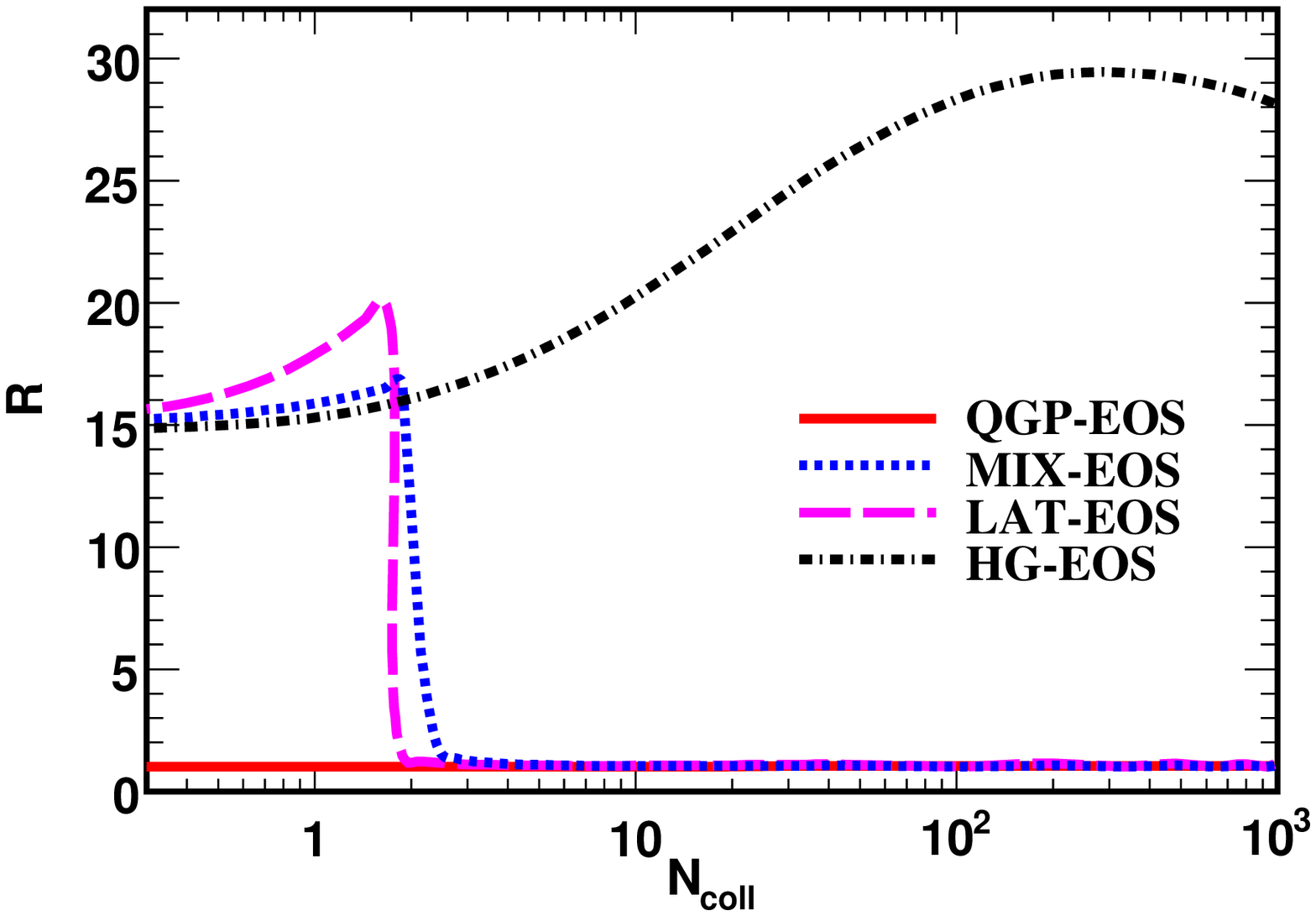}
\end{figure}

For non-central or peripheral collisions we can compute the elliptic
flow coefficient $v_{2}$ versus transverse momentum and invariant
mass, $v_{2}(p_T,M)=\left\langle \cos(2\phi_{p})\right\rangle $.
Setting the impact parameter $b=7$ fm, we calculat dilepton $v_2$ as
a function of invariant mass for four EOS, see the left panel of
Fig. \ref{fig:Mtscaling}. The results with the MIX-EOS or LAT-EOS
have similar features to Ref. \cite{Chatterjee:2007xk}. Dilepton
sources within different mass windows are sensitive to different
stages of the expansion history. We see that $v_2$ is much smaller
for larger $M$ than that for smaller $M$. This is because dileptons
at larger $M$ come mainly from the QGP phase at earlier time of the
fluid evolution with smaller collective flow, while those at smaller
$M$ are mainly from the HG phase at later stage when the collective
flow is fully developed. The HG contribution is characterized by the
peak at $\rho$ meson mass which quantitatively different from the
partonic contribution. If there is only one phase in the course of
the evolution, these distinct feature will disappear. 
The initial temperature of the QGP phase in EOS containing the QGP phase 
is much higher than that of the pure HG phase in HG-EOS at the 
same initial energy density. So the early stage emission from the former 
is much larger than that from the latter (since yield is proportional to $T^4$),
then the later stage emission for HG-EOS has much larger proportion 
in the total yield. This helps us understand the bigger elliptic
flow of HG-EOS than that of other EOS containing the QGP phase in the
intermediate mass region. As the temperature window for emission with 
HG-EOS is not very wide and the low mass dileptons come mostly
from the low temperature stage, similar to the surface emission of
hadrons, the elliptic flow has a slight increase with mass
in the low mass region. 
The peaks at the $\rho$ meson mass in the
MIX-EOS and LAT-EOS cases and the increase in the small mass region
are partially caused by the surface emission, but the main reason is
the influence of the QGP phase in the region below the $\rho$ meson
mass \cite{Chatterjee:2007xk}.

We also note that if the QGP is created, the dilepton spectrum in
the mass region between the $\phi$ meson mass and $J/\psi$ mass
depends approximately only on its transverse mass $M_T$. 
This can be seen from Eq. (\ref{eq:dndptdm02}), after the integration over azimuthal angle
$\phi _p$, the differential yield is proportional to
$I_0(\gamma_T v_T p_T/T)K_0(\gamma_T M_T/T)$. 
In the continuum region, the lepton mass, the quark mass and the
collectivity are negligible, then the yield is only a function of
$M_T$. This property is called the $M_T$ scaling \cite{Asakawa:1993kb}. 
It can be broken when transverse expansion becomes appreciable or hardonic gas
emission with extra mass scales dominate the yield.
By looking at the ratio,
\begin{equation}
R=\frac{dN/dM_T^2dp_T^2dy|_{M_T=2.6GeV,p_T=2GeV}}{dN/dM_T^2dp_T^2dy|_{M_T=2.6GeV,p_T=0GeV}},
\end{equation}
we study the $M_{T}$ scaling with more realistic hydrodynamic
evolution. As shown in the right panel of Fig. \ref{fig:Mtscaling},
the scaling is rather robust if the QGP phase is present. We find
that the elliptic expansion, the collective flow of the QGP and the
$\rho$ meson form factor do not qualitatively change the result in
Ref. \cite{Asakawa:1993kb}, if the di-electron yields from
partonic/hadronic source dominate in the intermediate/low mass
region respectively.

\section{summary and conclusion}

In conclusion, we have investigated the thermal production of
di-electrons in the quark-gluon plasma created in heavy ion
collisions at RHIC energy. Different types of the equations of state
are used in our study. The phase transition from hadron gas to
quark-gluon plasma leads to a rich structure for the thermal
dilepton production. The mass dependence of the inverse slope
parameter is sensitive to the collective dynamics of the medium: 
at the mass region ($M\lesssim $1 GeV), it is dominated by the hadronic interactions 
while in the intermediate mass region ($1\lesssim M\lesssim 3$ GeV), 
partonic interactions become important. 
At lower transverse momenta, the slope parameter for mixed-phase and lattice EOS  
shows different trends in $M$ in the low mass region (hadronic phase) 
and intermediate mass regions (partonic phase), which reflects 
the existence of two distinct phases. In this case, 
there is transition area around $M=1$ GeV to connect two $T_{eff}$ 
components with different trends.  
In the hadronic phase in the lower mass region the flow velocity is found to be
stronger than that in the partonic phase in the higher mass region.
Around $M\sim 1$ GeV, the slope parameter is found sensitive to the
equation of state of the fireball used in our calculation. The
elliptic flow and the $M_T$ scaling are calculated which show
distinct features for the hadronic and partonic phases. We have
investigated the surface emission or the emission in a fixed
interval of temperatures, where we have found that these features
are absent for the two phases. In this sense the dilepton emission
is a volume effect which can carry the information of the space-time
history of the fireball. Therefore all these features of the
collective flow can serve as clean probes to hot and dense medium
created in high energy heavy ion collisions.

Acknowledgement: JD and QW thank Hao-jie Xu for his help in 
adding the mesonic resonance contribution.  
QW is supported in part by the '100 talents'
project of Chinese Academy of Sciences (CAS) and by the National
Natural Science Foundation of China (NSFC) with grant No. 10735040.
NX is supported in part by the Department of Energy (DOE) with grant
No. DE-AC03-76SF00098. PZ is supported in part by the National
Natural Science Foundation of China (NSFC) with grant Nos. 10847001
and 10975084. JD is supported by China Postdoctoral Science
Foundation with grant No. 20080440722.

\end{document}